\documentclass{sig-alternate-05-2015}
\usepackage{multirow}
\usepackage{mathtools}
\usepackage{stmaryrd}
\usepackage{paralist}
\usepackage{url}
\usepackage{relsize}
\usepackage{multirow}
\usepackage{graphicx}
\usepackage{float}
\usepackage{multicol}
\usepackage{subfigure}
\usepackage{hyperref}
\usepackage{makeidx}  
\usepackage{graphicx}
\usepackage{pgfplots}
\usepackage{multirow}
\usepackage{amsmath}
\usepackage{amssymb}
\usepackage{subfigure}
\usepackage{ifsym}
\usepackage{stmaryrd}
\usepackage{algorithm}
\usepackage{algorithmic}
\usepackage{tikz,times}
\usepackage{xcolor}

\usepackage{lipsum}

\usepackage{array}
\usepackage{indentfirst}

\usepackage{tikz}
\usetikzlibrary{arrows,positioning,automata,decorations,fit,backgrounds,calc,shapes,snakes}
\usepgflibrary{shapes.geometric} 
\usetikzlibrary{shapes.geometric} 
\usepgflibrary{decorations.pathmorphing} 
\usetikzlibrary{decorations.pathmorphing} 
\usepackage{verbatim}
\usetikzlibrary{calc,backgrounds}
\usetikzlibrary{trees}

\begin{document}

\setcopyright{acmcopyright}

\doi{http://dx.doi.org/10.1145/0000000.0000000}

\isbn{978-1450317412}

\conferenceinfo{AOSD'12}{Hasso-Plattner Institut Potsdam, Germany,March 25--30, 2012}

\acmPrice{\$15.00}

%
\conferenceinfo{XXX}{XXX}
\CopyrightYear{2016} 
\crdata{0-12345-67-8/90/01}  

\title{MapSQ: A MapReduce-based Framework for SPARQL Queries on GPU}
%
%
%
%
%

\numberofauthors{3} 
%
\author{
%
%
Jiaying Feng$^{1,3}$ \hspace{0.1cm} Xiaowang Zhang$^{1,3}$ \hspace{0.1cm} Zhiyong Feng$^{2,3}$\\
       \affaddr{$^1$School of Computer Science and Technology,Tianjin University, Tianjin 300350, P.R.China}\\
       \affaddr{$^2$School of Computer Software,Tianjin University, Tianjin 300350, P. R. China}\\
       \affaddr{$^3$Tianjin Key Laboratory of Cognitive Computing and Application, Tianjin 300350, P.R.China}
       }
%

\maketitle
\begin{abstract}
In this paper, we present a MapReduce-based framework for evaluating SPARQL queries on GPU (named MapSQ) to large-scale RDF datesets efficiently by applying both high performance. Firstly, we develop a MapReduce-based Join algorithm to handle SPARQL queries in a parallel way. Secondly, we present a coprocessing strategy to manage the process of evaluating queries where CPU is used to assigns subqueries and GPU is used to compute the join of subqueries. Finally, we implement our proposed framework and evaluate our proposal by comparing with two popular and latest SPARQL query engines gStore and gStoreD on the LUBM benchmark.  The experiments demonstrate that our proposal MapSQ is highly efficient and effective (up to 50\% speedup).
\end{abstract}
%
%

%

%
%
%
%

\terms{Theory}


\section{Introduction}\label{sec:intro}
SPARQL is the standard query language over RDF datasets. With the rapid growth of RDF data and high complexity of SPARQL, processing RDF data effectively becomes a significant challenge. For instance, a SPARQL query as follow:\\
\textit{Q: SELECT ?person WHERE \{ ?person hasJob ?job. ?job workAt "Hospital".\}}

Many  approaches have been proposed to answer SPARQL queries, such as centralized engine gStore\cite{gStore}, and distributed engine\cite{gStoreD} Their efficiency depends on the calculative capabilities of CPU. However, the performance of CPU has almost reached its peak recently.

In recent years, graphic processing units (GPUs) has been adopted to process large-scale data, and their highly parallel structure makes them more efficient than general-purpose CPUs \cite{IRSMG}.  \cite{Mars} develops a MapReduce framework on GPU. MapReduce is a parallel programming model, which excels at large-scale data processing. Combining the characters of MapReduce and RDF datasets, we develop MapReduce-base join algorithm to improve the efficiency of RDF data processing.

SPARQL queries contains a series of triple patterns. Consider the following two triple patterns $P_1$(\textit{?person, hasJob, ?job}) and $P_2$(\textit{?job, workAt, "Hospital"}). Each triple pattern maps a partial matches set, shown as Table \ref{tab:tp1} and \ref{tab:tp2}. Both of them hold key and value. The key of them is their shared variable \textit{?job}, and the value is other variables. Final results can be obtained by joining  partial results, shown as Table \ref{tab:rs}.
\vspace*{-18pt}

\begin{table}[H]
\caption{Example of MapReduce-based Join}
\centering
\subtable[$Tp_1$]{
       \begin{tabular}{c|c}
       \hline
       \emph{Key1} & \emph{Value1} \\
        \hline
         Proffesor &Anny \\
         Doctor & Jim\\
         Nurse & Susan\\
        \hline
       \end{tabular}
       \label{tab:tp1}
}
\qquad
\subtable[$Tp_2$]{
       \begin{tabular}{c|c}
       \hline
       \emph{Key2} & \emph{Value2} \\
        \hline
        Doctor& "Hospital"\\
        Nurse& "Hospital"\\
        \hline
       \end{tabular}
       \label{tab:tp2}
}
\vspace*{-6pt}
\subtable[$RS$]{

       \begin{tabular}{c|c}
       \hline
       \emph{Key} & \emph{Value} \\
        \hline
        Doctor& Jim,"Hospital"\\
        Nurse& Susan,"Hospital"\\
        \hline
       \end{tabular}
       \label{tab:rs}
}
\end{table}
\vspace*{-8pt}
In this paper, we propose a SPARQL queries framework on GPU called MapSQ(MapReduce-based framework for SPARQL Queries), which implements results join based on MapReduce. Our work has three major contributions: (1) we propose a MapReduce-based join algorithm on GPU to match intermediate results, (2) we develop the MapSQ framework for improving SPARQL queries performance, and (3) extensive experiments confirm the efficiency and effectiveness of our method.

\section{MapSQ Framework}\label{sec:framework}

\textbf{MapReduce-based Join:} Join operation is the basic function in database system and has been researched extensively in CPU-based architecture. Many approaches have been designed by plain join algorithms, such as nested-loop join. We develop a novel method to improve the performance of centralized RDF engine.

\begin{algorithm}[htb]
\caption{MapReduce-based Join}
\label{alg:Framwork}
\begin{algorithmic}[1]
\REQUIRE $Tp_1$ and $Tp_2$ are the partial matches of each triple pattern.
\ENSURE  $RS$, the join results of two triple patterns.
\STATE \textbf{function} Map($Tp{_1}$, $Tp_2$)
\FOR {each item $\in$ ($Tp{_1}$, $Tp_2$)}
 \STATE split and emit intermediate.
\ENDFOR
\STATE sort intermediates.
\STATE \textbf{function} ReduceDuplicate
\FOR {each key}
 \IF{value1's label is RIGHT and value2's label is LEFT}
  \STATE put the key-value pair into $RS$.
 \ENDIF
\ENDFOR
\RETURN $RS$;
\end{algorithmic}
\end{algorithm}

In our work, join operation merges partial results after each triple patterns is matched.
The output of matching is delivered to join if triple patterns contain one shared variables.

Algorithm \ref{alg:Framwork} shows the MapReduce-based join on GPU.
In order to make full use of parallelization advantages, the join operation divided into three phases: Map, Sort and Reduce duplicate.
When mapping phase, we set flag according to the shared variable's position in triples and split each triple.
The flag, including left or right, contributes to reduce unnecessary computation in the stage of reducing duplicate.
GPU's Single Instruction Multiple Data architectures contribute to accelerate cartesian product in parallel.

\begin{figure}[H]
\centering
\includegraphics[width=0.4\textwidth]{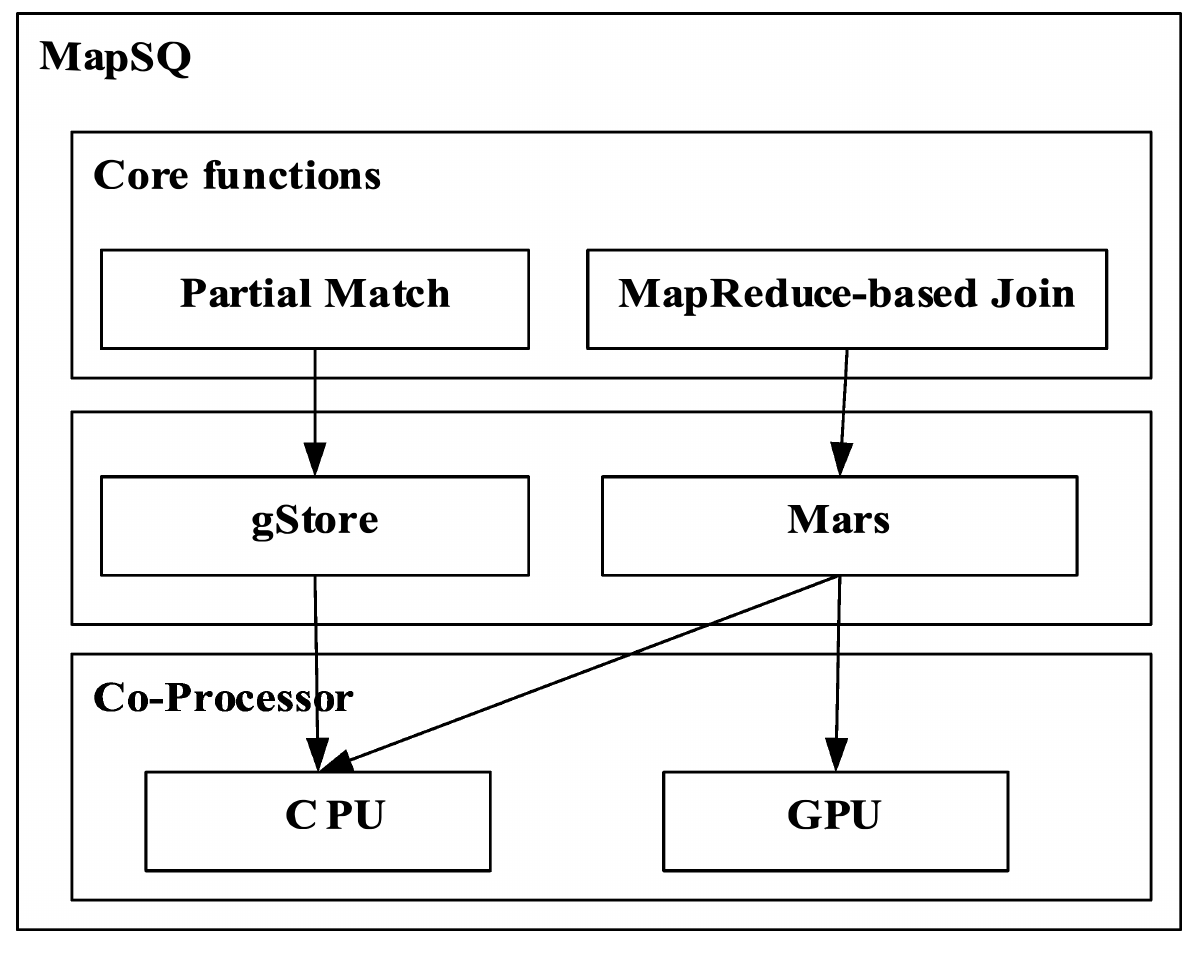}
\caption{MapSQ Framework}\label{fig:MapSQ}
\end{figure}

\textbf{MapSQ Framework:} We present design and implementation of MapSQ framework, shown in Figure \ref{fig:MapSQ}. The bottom of framework is GPU and CPU-based gStore, which is used as a black box for answering SPARQL queries.
\vspace*{-4pt}
\vspace*{-1pt}

When answering SPARQL queries, there are two steps to process queries: partial matching and MapReduce-based join.
The first step is each triple pattern matches the partial results  through centralized RDF engine in parallel, which adopts gStore to dispose queries.
The second step is computing the final results by assembling partial matchings. MapReduce-based join Algorithm is implemented on GPU.

\vspace*{-8pt}
\section{Experiments  and EVALUATIONS}\label{sec:experiments}
In this section, we present the performance of MapSQ framework in comparison with CPU-based methods, including gStore and gStoreD. The purpose of the experiments is to evaluate the improvement of efficency. 
Our experiments were carried out on a PC with a GTX590 GPU and an Intel(R) Core(TM) i7-2600 Quad processor running Linux ubuntu14.04.
The CPU main memory is 2GB, and the GPU device memory is 1536MB.


We used LUBM \footnote{http://swat.cse.lehigh.edu/projects/lubm/} as the benchmark datasets in our experiments. LUBM is an ontology for university domain and generates arbitrary scale dataset.
We choose 5 benchmark queries to test queries performance. gStore and gStoreD are better performance methods in centralized RDF engine.


\begin{table}[h]
\centering
\caption{join time(ms)}\label{tab:join}
\begin{tabular}{c|c|c|c|c|c}
\hline
\emph{Query}&\emph{gStore}&\emph{gStoreD} &\emph{MapQS} &\emph{$SpeedUp_g$} &\emph{$SpeedUp_D$}\\
\hline
Q1 &1677 &2079 &1429 &1.17 &1.45\\
Q2 &3886 &4131 &3081 &1.26 &1.34\\
Q3 &433  &442 &351  &1.23 &1.25\\
Q4 &635  &698 &339 &1.87 &2.05\\
Q5 &7254 &10875 &6292 &1.15 & 1.72\\
\hline
\end{tabular}
\end{table}


%
%

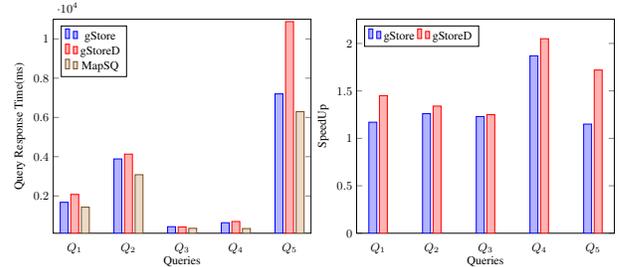
\begin{figure}[h]
\subfigure[Queries Response Time] {
\begin{minipage}[t]{0.45\linewidth}
\centering
\scalebox{0.5}{
\begin{tikzpicture}
\begin{axis}[
 major x tick style = transparent,
 ybar,
 ymax = 11000,
 ymin = 100,
 ylabel = {Query Response Time(ms)},
 xlabel = {Queries},
 symbolic x coords = {$Q_1$,$Q_2$,$Q_3$,$Q_4$,$Q_5$},
 scaled y ticks = true,
 ylabel near ticks,
 xlabel near ticks,
 bar width=6pt,
 xlabel style={below=-0.15cm},
     legend pos= north west
 ]

\addplot coordinates {($Q_1$, 1677) ($Q_2$, 3886) ($Q_3$, 433) ($Q_4$, 635) ($Q_5$, 7200)};
\addplot coordinates {($Q_1$, 2079) ($Q_2$, 4131) ($Q_3$, 422) ($Q_4$, 698) ($Q_5$, 10879)};
\addplot coordinates {($Q_1$, 1429) ($Q_2$, 3081) ($Q_3$, 351) ($Q_4$, 339) ($Q_5$, 6292)};
 \legend{gStore,gStoreD,MapSQ}
\end{axis}
\end{tikzpicture}
}
\end{minipage}
}
\subfigure[SpeedUp] {
\begin{minipage}[t]{0.45\linewidth}
\centering
\scalebox{0.5}{
\begin{tikzpicture}
\begin{axis}[
 major x tick style = transparent,
 ybar,
 ymin = 0,
 ylabel = {SpeedUp},
 xlabel = {Queries},
 symbolic x coords = {$Q_1$,$Q_2$,$Q_3$,$Q_4$,$Q_5$},
 scaled y ticks = true,
 ylabel near ticks,
 xlabel near ticks,
 bar width=6pt,
 xlabel style={below=-0.15cm},
    legend columns=-1, 
     legend pos= north west
 ]
\addplot coordinates {($Q_1$, 1.17) ($Q_2$, 1.26) ($Q_3$, 1.23) ($Q_4$, 1.87) ($Q_5$, 1.15)};
\addplot coordinates {($Q_1$, 1.45) ($Q_2$, 1.34) ($Q_3$, 1.25) ($Q_4$, 2.05) ($Q_5$, 1.72)};
 \legend{gStore,gStoreD}
\end{axis}
\end{tikzpicture}
}
\end{minipage}
}
\caption{Query performance}\label{tik:query}
\end{figure}

Table\ref{tab:join} and Figure \ref{tik:query} show the query response time of gStore, gStoreD and MapSQ. And the speedup is computed among them. Our MapReduce-based join algorithm is faster than the join operation of gStore and gStoreD. Comparing with gStore and gStoreD, MapSQ queries response time reduce 15\%-50\%, shown in Figure \ref{tik:query}(a). the spreedup among them shown in Figure \ref{tik:query}(b).
Experiment results show that our framework performs better than the algorithms on CPU and can efficiently speed up SPARQL queries answering.

Experiment results show that our framework performs better than the algorithms on CPU and can efficiently speed up SPARQL queries answering. Furthermore, there has been a significant increase in different
queries. For instance, we observe $Q_4$, which reduce to the half of CPU-based engine. In this scene, MapSQ shortens query response time especially large dataset scale.

\vspace*{-8pt}

\section{Conclusions}\label{sec:conclusion}
In this paper, we proposed a MapReduce-based framework for SPARQL queries on GPU and the preliminary experiments show our proposal can evaluate SPAQRL queries on large-scale RDF datasets effectively and effiently (up to 50\%, without further optimization). We think that our proposal will bring new methods in processing large-scale RDF data by integrating parallel architecture with new hardwares. This work is supported by the programs of the National Key Research and Development Program of China (2016YFB1000603) and the National Natural Science Foundation of China (61672377).


\vspace*{-8pt}
\begin{thebibliography}{5}


\bibitem{Mars}
B.~He, W.~Fang, Q.~Luo, N.~Govindaraju, and T.~Wang.
\newblock   Mars: a MapReduce framework on graphics processors
\newblock In: {\em Proc. of {PACT}'08}, pp. 260-269, 2008.

\bibitem{gStoreD}
P.~Peng, L.~zou, MT.~Zsu, L.~Chen, and D.~Zhou.
\newblock  Processing SPARQL queries over distributed RDF graphs.
\newblock {\em J. VLDB.}, 25(2): 243-268, 2016.



\bibitem{gStore}
L.~Zou, J.~Mo, L.~Chen, MT.~Zsu and D.~Zhou.
\newblock  gStore: Answering SPARQL queries via subgraph matching.
\newblock In: {\em Proc. of {VLDB Endowment}}, pp.482-493, 2011.

\bibitem{IRSMG}
J.~Zhang, B.~Zhang, X.~Zhang, and Z.~Feng.
\newblock   IRSMG: Accelerating Inexact RDF Subgraph Matching on the GPU
\newblock In: {\em Proc. of {ISWC Poster}}, 2016.

\bibitem{Relational}
B.~He, K.~Yang, R.~Fang, M.~Lu N.~Govindaraju, Q.~Luo, and P.Sander.
\newblock Relational joins on graphics processors
\newblock In: {\em Proc. of {ISIGMOD}'08}, pp. 511-524, 2008.



\end{thebibliography}

\end{document}